\documentclass[preprint,12pt]{aastex}

\usepackage{emulateapj5}

\slugcomment{Accepted for Publication in ApJ}

\shorttitle{On S Ori 70}
\shortauthors{Burgasser et al.}

\begin{document}

\title{S Ori 70: Just a Foreground Field Brown Dwarf?}

\author{
Adam J.\ Burgasser\altaffilmark{1,2},
J.\ Davy Kirkpatrick\altaffilmark{3},
Mark R.\ McGovern\altaffilmark{1},
Ian S.\ McLean\altaffilmark{1},
L.\ Prato\altaffilmark{1},
and I.\ Neill Reid\altaffilmark{4},
}

\altaffiltext{1}{Department of Physics \& Astronomy,
University of California
at Los Angeles, Los Angeles,
CA, 90095-1562; adam@astro.ucla.edu, mclean@astro.ucla.edu, mcgovern@astro.ucla.edu,
lprato@astro.ucla.edu}
\altaffiltext{2}{Hubble Fellow}
\altaffiltext{3}{Infrared Processing and Analysis Center, M/S 100-22,
California Institute of Technology, Pasadena, CA 91125; davy@ipac.caltech.edu}
\altaffiltext{4}{Space Telescope Science Institute, 3700 San
Martin Drive, Baltimore, MD 21218; inr@stsci.edu}

\begin{abstract}
We examine recent claims that the
T-type brown dwarf S Ori 053810.1-023626 (S Ori 70)
is a spectroscopically verified low mass (3$^{+5}_{-1}$ M$_{Jup}$)
member of the 1--8 Myr $\sigma$ Orionis cluster.
Comparative arguments by Mart{\'{i}}n \& Zapatero Osorio
asserting that S Ori 70 exhibits low surface gravity
spectral features indicative of youth and low mass
are invalidated by the fact that their
comparison object was not the field T dwarf 2MASS 0559$-$1404 but rather
a nearby background star.  Instead, we find that the
1--2.5 $\micron$ spectra of S Ori 70 are
well-matched to older (age $\sim$ few Gyr)
field T6--T7 dwarfs.  Moreover, we find that spectral model fits to
late-type field T dwarf spectra tend to
yield low surface gravities
($\log{g}$ = 3.0--3.5), and thus young ages ($\lesssim$ 5 Myr) and
low masses ($\lesssim$ 3 M$_{Jup}$), inconsistent
with expected and/or empirical values.
Finally, we show that the identification of one
T dwarf in the field imaged by Zapatero Osorio et al.\ is
statistically consistent with the expected foreground contamination.
Based on the re-examined evidence, we conclude
that S Ori 70 may simply be an old, massive (30--60 M$_{Jup}$) field
brown dwarf lying in the foreground of the $\sigma$ Orionis cluster.
This interpretation should be considered before presuming the existence of
so-called ``cluster planets.''
\end{abstract}


\section{Introduction}

The boundary between classical planets (e.g., Jupiter) and
stars has been blurred by the discovery of brown dwarfs, objects which
presumably form like stars but lack sufficient mass to sustain core
Hydrogen fusion \citep{hay63,kum63}.  Known brown dwarfs exhibit a number
of features similar to those of giant planets, including size\footnote{We
adopt the Jupiter-based units
R$_{Jup}$ = 7.15$\times$10$^9$ cm and
M$_{Jup}$ = 1.90$\times$10$^{30}$ g = 0.00095 M$_{\sun}$ throughout this article
\citep{tho00}.}
(R $\sim$ R$_{Jup}$),
atmospheric composition (e.g., CH$_4$,
condensate clouds), and possibly mass (of order 10 M$_{Jup}$); but are
typically found in stellar environments -- as isolated field
and cluster objects and as wide companions to main sequence stars.
While brown dwarfs appear to be rare as close companions to stars \citep{mar00},
a regime occupied by Solar and extrasolar planets,
the possibility of planetary ejection \citep{li02,nel03}
or brown dwarf capture \citep{bon03}, and the identification of
low mass ratio binaries with substellar secondaries
(e.g., HR 7329AB; Lowrance et al.\ 2000),
makes it impossible to unambiguously deduce
the origin of a particular substellar object.
This situation has led to considerable debate over the definition of
a planet and its distinction from a low-mass brown dwarf
\citep{bos03}.

Recently, \citet[hereafter Z02]{zap02} have identified a brown dwarf in
the direction of $\sigma$ Orionis
particularly relevant to this debate.  The object,
S Ori 053810.1-023626 (hereafter S Ori 70), is classified as a T dwarf based on
the presence of CH$_4$ absorption in its near-infrared
spectrum \citep{me02a,geb02}, and therefore has an effective temperature (T$_{eff}$)
less than $\sim$1200--1400 K \citep{kir00,leg01,stp01}.
Z02 identify this object as a candidate member of the young
(age $\sim$ 1-8 Myr) $\sigma$ Orionis cluster, and deduce an estimated mass of only
3$^{+5}_{-1}$ M$_{Jup}$, potentially the
least massive brown dwarf known.
Both Z02 and \citet[hereafter MZ03]{mar03}
claim that cluster membership for S Ori 70 is verified by
(1) comparative near-infrared spectroscopy with the presumably older (higher surface gravity)
field T dwarf 2MASS J05591914$-$1404488 \citep{me00c};
(2) spectral fits to theoretical models from \citet{all01}
that yield a low surface gravity
consistent with a young, low mass brown dwarf; and
(3) low foreground contamination
by field T dwarfs in the search area imaged by Z02.

In this article, we address flaws in each of these
arguments.  Specifically, in ${\S}{\S}$2--4 we show that (1) comparison with the spectrum of
2MASS 0559$-$1404 is invalidated by the fact
that MZ03 observed the wrong comparison star, and in fact the spectrum
of S Ori 70 is an excellent match
to those of field T6--T7 dwarfs; (2) \citet{all01}
spectral model fits tend to be skewed toward lower surface gravities, and thus lower
ages and masses, for late-type T dwarfs; and (3) the identification of
one foreground T dwarf in the Z02 field is statistically
consistent with the expected
contamination.  This analysis indicates that S Ori 70 may
simply be a foreground field T dwarf with a much older age and
higher mass than reported by Z02 and MZ03.
We summarize our results in $\S$5.

\section{Comparison of the J-band Spectrum of S Ori 70 to Field Objects}

Young, low-mass brown dwarfs have much lower surface gravities than
their evolved field counterparts; $\log{g}$(cgs) $\sim$ 3--4 is typical
for a 1--10 Myr, 1--50 M$_{Jup}$ object, versus 4.5--5.5 for a
$\sim$1 Gyr field brown dwarf with similar T$_{eff}$ \citep{bur97,bar03}.
As the abundances and band strengths of molecular
species that dominate the photospheric opacity of cool brown dwarfs
(e.g., H$_2$O, CH$_4$, collision-induced H$_2$ absorption)
are pressure-sensitive, the photospheric gas pressure, and hence surface gravity,
can substantially influence the emergent spectral energy distribution.

The $J$-band spectrum of S Ori 70 presented in MZ03 exhibits a distinct triangular-shaped
spectral morphology which these authors argue is due to its
low surface gravity. They draw an
analogy with the peaked spectral morphologies of low-gravity L dwarfs
in the Trapezium cluster \citep{luc01}, although the latter data were
obtained at $H$- and $K$-bands and not at $J$.  The low surface gravity
argument of MZ03 is supported by a comparison with the spectrum of
the presumably older, and hence higher surface gravity, field T dwarf 2MASS 0559$-$1404,
which appears in their Figure 1 to have a very different $J$-band spectral morphology.
However, their reported spectrum is not that of
2MASS 0559$-$1404, but rather an adjacent background source.
Figure 1 compares the reported MZ03 spectrum of 2MASS 0559$-$1404 to data
obtained as part of the NIRSPEC Brown Dwarf Spectroscopic
Survey \citep[hereafter BDSS]{mcl03}; and to a re-extraction of the raw
data from MZ03 using the REDSPEC package\footnote{See
\url{http://www2.keck.hawaii.edu/inst/nirspec/redspec/index.html}.
Data reduction procedures are fully described in \citet{mcl03}.} (Figure 1).
The re-reduced spectrum of the comparison
object observed by MZ03 exhibits none of the hallmark features
of mid- and late-type T dwarfs.
Thus, the differences noted by MZ03 between S Ori 70 and their field comparison star
have nothing to do with surface gravity effects.

In fact, the triangular shape of the $J$-band spectrum of S Ori 70 is
common amongst mid- and late-type T dwarfs, the result of
H$_2$O and CH$_4$ absorption wings on either side of the $J$-band peak \citep{me02a,geb02}.
Figure 2 compares the MZ03 spectrum of S Ori 70 to BDSS data for
the T7 field brown dwarf 2MASS J15530228+1532369 \citep{me02a}.
This object is presumably much older than the $<$10 Myr $\sigma$ Orionis cluster
given the similarity of its 1--2.5 $\micron$
spectrum to other T6--T7 field dwarfs \citep{me02a,mcl03}, which have space motions
consistent with 1--5 Gyr disk dwarfs \citep{giz00,dah02,tin03}, and the absence of any
star forming regions in its vicinity.
The match between the S Ori 70 and 2MASS 1553+1532 data is excellent, and
similar good agreement is found with other
T6 and T7 BDSS spectra.  Hence,
the $J$-band spectrum of S Ori 70 is consistent with that of an
old, high-gravity, late-type field T dwarf.

\section{Spectral Model Fits: Skewed Ages for Mid-type T dwarfs}

Both Z02 and MZ03 determine the surface gravity and T$_{eff}$ for S Ori 70
by fitting their spectra to
the most recent theoretical models from \citet{all01}.  In the absence
of dynamical mass, radius, and bolometric flux measurements, the use
of spectral models is required to derive these physical parameters, as the
complexity of the spectra prohibit classical techniques such as curve of growth
\citep{pav96}.
However, despite substantial improvements over
the past decade, these models do not yield reliable fits to the observed data,
particularly when disentangling T$_{eff}$ and gravity \citep{sau00,geb01}.
These limitations are
largely due to incomplete or inaccurate molecular line lists and
the uncertain treatment of photospheric condensates \citep{all01,me02c}.
For instance, \citet{leg01} find T$_{eff}$ discrepancies as high as 400 K between
spectral model fits and luminosity-determined temperatures for the latest-type
field L dwarfs.

To assess whether similar discrepancies occur in the T dwarf regime,
we performed minimum ${\chi}^2$ fits of BDSS $J$-band data for seven
T6--T7 field dwarfs
using the same \citet{all01} COND models as those used by MZ03.
We sampled a grid of models spanning
$2.5 < \log{g} < 6.0$ and 600 $<$ T$_{eff}$ $<$ 1400 K in intervals of 0.5 dex and 100 K,
respectively.  Both empirical and model spectra were Gaussian smoothed to the
instrumental resolution (R $\sim$ 2000) and interpolated onto
identical wavelength scales.  As parallax measurements are unavailable
for most of these objects and S Ori 70, we performed the fits
by normalizing the empirical and theoretical spectra at the 1.27 $\micron$ spectral peak.
The ${\chi}^2$ deviation between the spectra
was computed over the range 1.16--1.34 $\micron$, minimizing over
relative scalings of 0.8--1.2.  This procedure is similar to that employed by MZ03
for S Ori 70.

Table 1 lists the best-fit model parameters for the spectral data,
along with associated mass and age estimates from the \citet{bar03}
evolutionary models.  Given that the objects examined in these fits are
all field dwarfs, it is
readily apparent that the derived parameters tend to be
skewed toward lower gravities,
and hence younger ages and lower masses, particularly for the T7 dwarfs.
The most deviant case is that of Gliese 570D, a widely-separated T8
brown dwarf companion to the nearby Gliese 570 triple star system
\citep{me00a}.  Assuming coevality, Gliese 570D has a well-defined age of
2--5 Gyr \citep{geb01}.
On the contrary, the spectral model
fits predict an age $<$ 1 Myr and mass $<$ 1 M$_{Jup}$.
Figure 3 compares the minimum ${\chi}^2$ fit model for this object to one
based on the more widely-adopted parameters
$\log{g} = 5.0$ and T$_{eff}$ = 800 K \citep{geb01}.
The lower gravity model matches the overall J-band morphology better
than the higher-gravity model,
but predicts spiked features around 1.255 $\micron$.  These features
do not coincide with (presumably noise) spikes seen at 1.265 $\micron$
in the S Ori 70 spectrum, as is evident in
Figure 3 of MZ03.  The inferior fit of the high gravity model is generally confined
to the spectral peak, and may be the result
of deep condensate opacity in this relatively transparent spectral region
\citep{bur02}.  Regardless of the reason, it is clear that
the best fit spectral models can yield skewed gravities for late-type field T dwarfs,
resulting in underestimated ages and masses.  Hence, the spectral fit parameters
for S Ori 70, which itself appears to be a late-type T dwarf, must be considered
with skepticism.

There are also inconsistencies between the $J$-band spectral model fits
for S Ori 70 from MZ03 and the $HK$ spectral model fits
from Z02.  The latter yield a substantially
lower T$_{eff}$ (800 versus 1100 K) and a somewhat higher surface gravity
(4.0 versus 3.5 dex), which is in fact consistent with a 50 Myr, not 8 Myr, brown dwarf
\citep{bar03}.  While the differences between the parameter fits
are within the reported uncertainties, they yield significantly incompatible spectral
morphologies.  Figure 4 demonstrates this,
comparing the $HK$ spectrum from Z02 to a smoothed
\citet{all01} model using the best fit parameters from MZ03.
It is clear that this model grossly underestimates
the strength of the 1.6 $\micron$ CH$_4$ band and somewhat overestimates
the flux at $K$-band.  On the other hand, empirical spectral data for the field T6.5
2MASS J10475385+2124234
\citep{me99,me02a}, obtained with identical instrument settings, does fairly well at
matching the $H$-band region, although it somewhat underestimates the flux at $K$-band.
The latter disagreement, possibly due to differences in collision-induced H$_2$ opacity,
may be indicative of a somewhat lower surface gravity for S Ori 70 relative to
2MASS 1047+2124 \citep{leg03}, although signal-to-noise may also be an issue.
Nevertheless, the fact that
empirical data for field T6--T7 dwarfs reproduce the 1.1--1.3 $\micron$
and 1.5--2.4 $\micron$
spectra for S Ori 70 more consistently than the theoretical models
indicates both the unreliability of physical parameters derived from the model fits
and the likelihood that S Ori 70 is a foreground brown dwarf.

\section{The Statistical Significance of Foreground Contamination}

A third argument used by Z02 to establish cluster membership for S Ori 70 is the
low foreground contamination by field T dwarfs (0.08--0.3 objects)
expected in their 55.4 arcmin$^2$
search area.  We have
computed an independent estimate of the foreground contamination
using substellar luminosity function simulations from
Burgasser (2001); absolute photometry from \citet{dah02} and \citet{tin03}; and
an assumed limiting magnitude of $J \sim 21$ (Z02).
For a substellar mass function $dN/dM \propto M^{-\alpha}$
and $0.5 \leq \alpha \leq 1.5$, we find 0.1--0.2 T dwarfs
with 700 $\leq$ T$_{eff}$ $\leq$ 1100 K (typical for late-type T dwarfs; Burgasser 2001)
expected in the Z02 field, consistent
with their estimates and certainly less than unity.

However, the relevant quantity is the likelihood of detecting
one foreground object in the imaged area given
the expected contamination rate.  This confidence limit, CL, can be derived from
Poisson statistics:
\begin{equation}
CL = \sum_{x=0}^{n-1} \frac{{\lambda}^xe^{-{\lambda}}}{x!} = e^{-{\lambda}}
\end{equation}
\citep{geh86} for $n = 1$ source detected
and an expected contamination of $\lambda$.
For $0.08 < \lambda < 0.3$, Eqn.\ 1 yields $0.74 < CL < 0.92$, equivalent
to 0.6--1.4 $\sigma$ on a Gaussian scale.  Hence, the presence of one foreground T dwarf
in the Z02 field is a 1$\sigma$ event, and cannot be ruled out statistically.

\section{Discussion}

Based on the arguments above, we find compelling evidence that
S Ori 70 is not a member of the $\sigma$ Orionis cluster but rather
a foreground field brown dwarf.
This is not an unexpected result given the predicted $\sim$30\% contamination rate
amongst L dwarfs in the $\sigma$ Orionis sample \citep{zap00}.
Indeed, quite a few $\sigma$ Orionis
brown dwarf candidates have been ruled
out as foreground objects in follow-up observations
\citep{bej01,mar01,ken02,zap02b,brd03,muz03,mcg04},
and \citet{ken02} find the fraction of spectroscopically-verified candidates
in $\sigma$ Orionis decreases toward fainter magnitudes.
It should come as no surprise that the faintest candidate to date, S Ori 70,
may be a contaminant field source.

Uncertainties in the spectral models and difficulty in obtaining high
signal-to-noise spectra for this faint T dwarf ($J = 20.28{\pm}0.10$; Z02)
imply that
unambiguous verification of cluster membership for S Ori 70 will require
measurement of its parallax and proper motion to establish spatial
and kinematic association.  Such
observations are hindered by the
distance of this object even if it is a foreground brown dwarf.
Z02 measure an upper limit proper motion
of $\mu \leq 0{\farcs}1$ yr$^{-1}$ for S Ori 70,
which only restricts its distance to $\gtrsim$ 40 pc,
roughly half the spectrophotometric distance if it is a field T6--T7 dwarf ($\sim$ 75--100 pc;
Tinney, Burgasser, \& Kirkpatrick 2003).

We conclude that S Ori 70 has not been rigorously proven
to be a member of the young $\sigma$ Orionis
cluster.  Rather, observations obtained thus far are consistent with this object
being a massive (M $\sim$ 30--60 M$_{Jup}$ assuming an age of 1--5 Gyr)
foreground field brown dwarf.  This interpretation should be seriously considered
before assuming the existence of a ``cluster planet'' population
in the $\sigma$ Orionis cluster.

\acknowledgments

The authors would like to thank E.\ L.\ Mart{\'{i}}n and M.\ R.\ Zapatero Osorio
for candid discussions regarding their results and for providing their reduced and raw
data for S Ori 70 and 2MASS 0559$-$1404; F.\ Allard for providing
access to theoretical spectral models; and
T.\ Ayres and L.\ Hillenbrand for very useful discussions.
We also thank our anonymous referee for her/his very prompt review.
A.\ J.\ B.\ acknowledges support provided by NASA through
Hubble Fellowship grant HST-HF-01137.01 awarded by the Space Telescope Science Institute,
which is operated by the Association of Universities for Research in Astronomy,
Incorporated, under NASA contract NAS5-26555.
This research has made use of
the SIMBAD database, operated at CDS, Strasbourg, France.

\clearpage

\begin{deluxetable}{llcccccccc}
\tabletypesize{\tiny}
\tablecaption{Comparison of Model Fits to Expected Parameters for
 Field T6--T8 Dwarfs}
\tablewidth{0pt}
\tablehead{
 & &
\multicolumn{4}{c}{Model Fit Parameters\tablenotemark{b}} & &
\multicolumn{3}{c}{Expected Field Parameters\tablenotemark{c}} \\
\cline{3-6} \cline{8-10}
\colhead{Object} &
\colhead{SpT\tablenotemark{a}} &
\colhead{T$_{eff}$ (K)} &
\colhead{$\log{g}$ (cgs)} &
\colhead{Age (Gyr)} &
\colhead{Mass (M$_{\sun}$)} & &
\colhead{T$_{eff}$ (K)} &
\colhead{$\log{g}$ (cgs)} &
\colhead{Mass (M$_{\sun}$)} \\
\colhead{(1)} &
\colhead{(2)} &
\colhead{(3)} &
\colhead{(4)} &
\colhead{(5)} &
\colhead{(6)} & &
\colhead{(7)} &
\colhead{(8)} &
\colhead{(9)} \\
}
\startdata
2MASS J23565477+1553111 & T6 & 1100 & 5.0 & 0.8 & 0.03 &  & 1000 & 5.0 & 0.03--0.06 \\
SDSSp J162414.37+002915.6 & T6 & 900 & 5.0 & 2 & 0.03 & & 1000 & 5.0 & 0.03--0.06 \\
2MASS J12373919+6526148 & T6.5 & 900 & 5.0 & 2 & 0.03 &  & 950 & 5.0 & 0.03--0.05 \\
2MASS J15530228+1532369 & T7 & 1200 & 3.5 & 0.004 & 0.003 &  & 900 & 5.0 & 0.03--0.05 \\
2MASS J07271824+1710012 & T7 & 1200 & 3.5 & 0.004 & 0.003 & & 900 & 5.0 & 0.03--0.05 \\
Gliese 570D & T8 & 1100 & 3.0 & $<$ 0.001 & $<$ 0.001 & & 784--824\tablenotemark{d} & 5.0--5.3\tablenotemark{d} & 0.03--0.05\tablenotemark{d} \\
2MASS J04151954$-$0935066 & T8 & 700 & 4.5 & 0.7 & 0.014 & & 800 & 4.5--5.0 & 0.02--0.04 \\
\enddata
\tablenotetext{a}{Near-infrared spectral types from \citet{me02a}.}
\tablenotetext{b}{T$_{eff}$ and $\log{g}$(cgs) for the best (lowest ${\chi}^2$)
\citet{all01} model fits to BDSS $J$-band data;
ages and masses are derived from the evolutionary models
of \citet{bar03}.}
\tablenotetext{c}{T$_{eff}$ estimated from the empirical relation
T$_{eff} \approx 1600 - 100{\times}$SpT, where SpT(T0) = 0, SpT(T5) = 5, etc \citep{me01};
$\log{g}$(cgs) and masses from \citet{bar03} assuming an age of 1--5 Gyr.}
\tablenotetext{d}{Empirical parameters from \citet{geb01} assuming an age of
2--5 Gyr, estimated from the system's K4 V primary star.}
\end{deluxetable}

\clearpage

\begin{figure}
\centering
\plotone{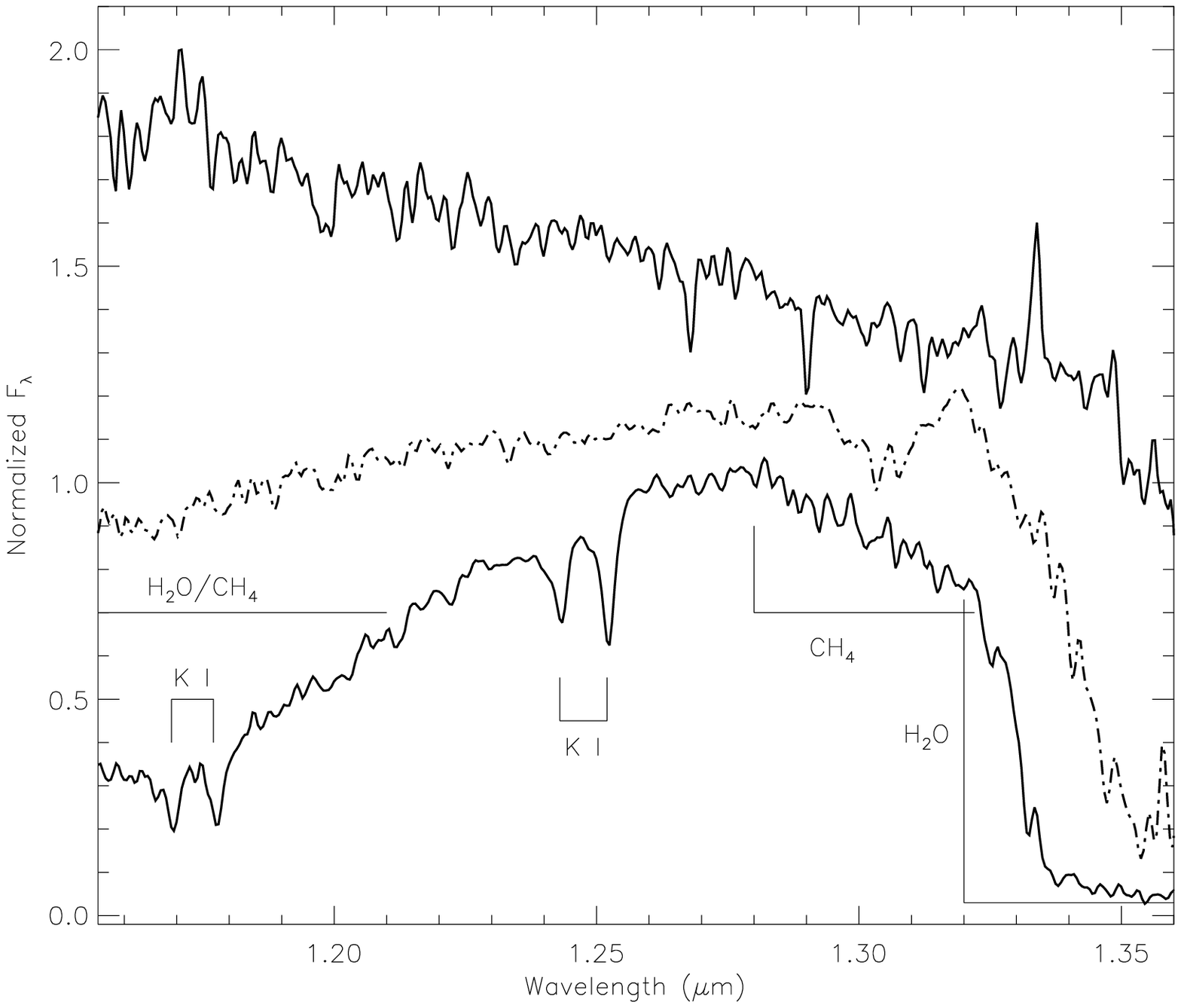}
\caption{Comparison of the reported $J$-band spectrum of 2MASS 0559$-$1404 from
MZ03 (middle spectrum, dot-dashed line) to BDSS data for the same object
(bottom spectrum)
and our re-reduction of the raw data from MZ03 using REDSPEC (top spectrum).
All three spectra are normalized at 1.27 $\micron$ and offset by a constant for clarity.
The object observed by MZ03 is a background star, and the features
present in their reported spectrum are likely due to a saturated flat field.
None of the indicated features typical of T dwarf spectra are seen in the MZ03 data.}
\end{figure}

\begin{figure}
\centering
\plotone{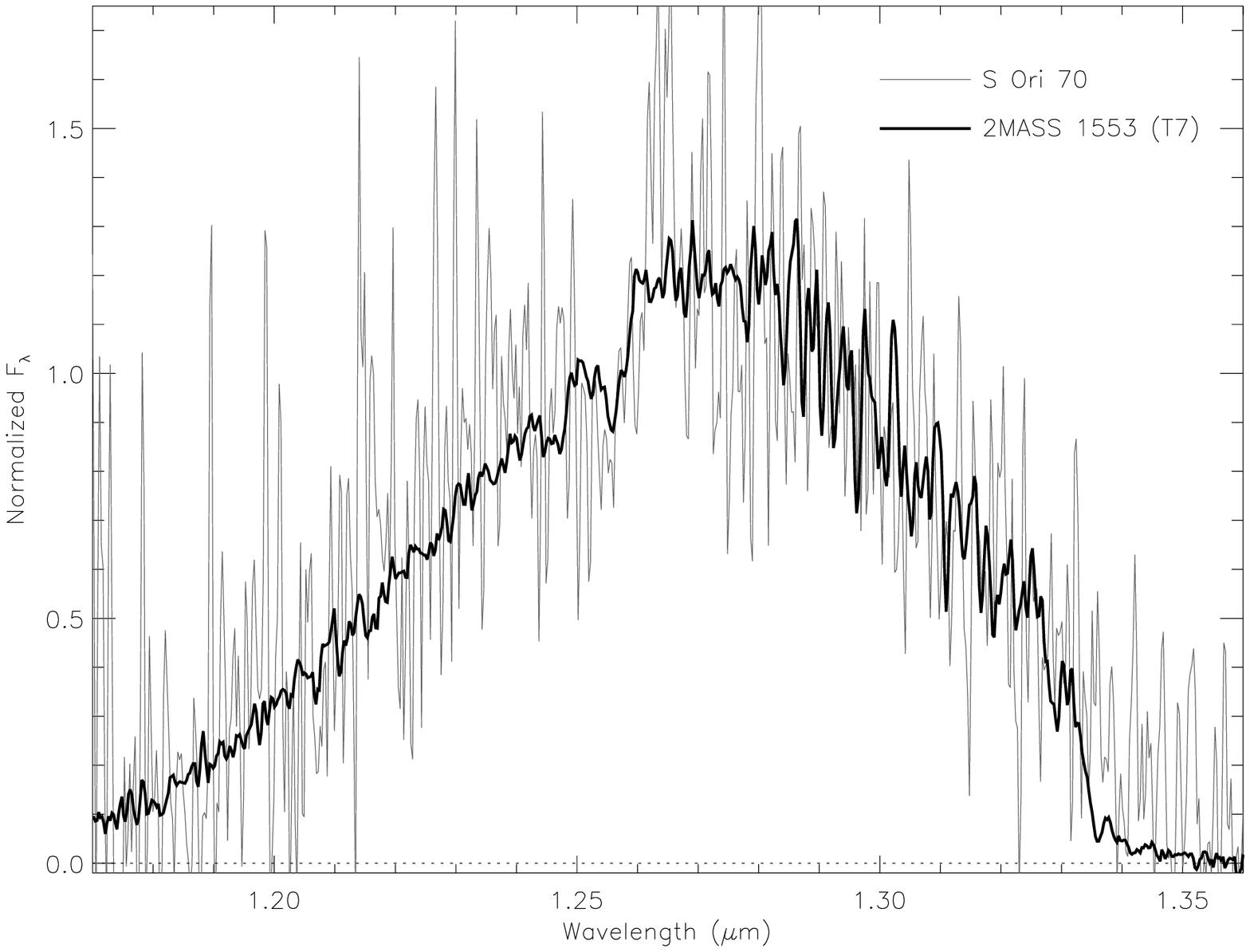}
\caption{Comparison of the $J$-band spectrum of S Ori 70 from MZ03 (black line)
to NIRSPEC BDSS data for the field T7 brown dwarf
2MASS 1553+1532 (black line).
Data are normalized at 1.27 $\micron$.
The excellent match between these spectra and other presumably old and therefore
high gravity field T6--T7 dwarfs
belie arguments that S Ori 70 exhibits low-gravity signatures.}
\end{figure}

\begin{figure}
\centering
\plotone{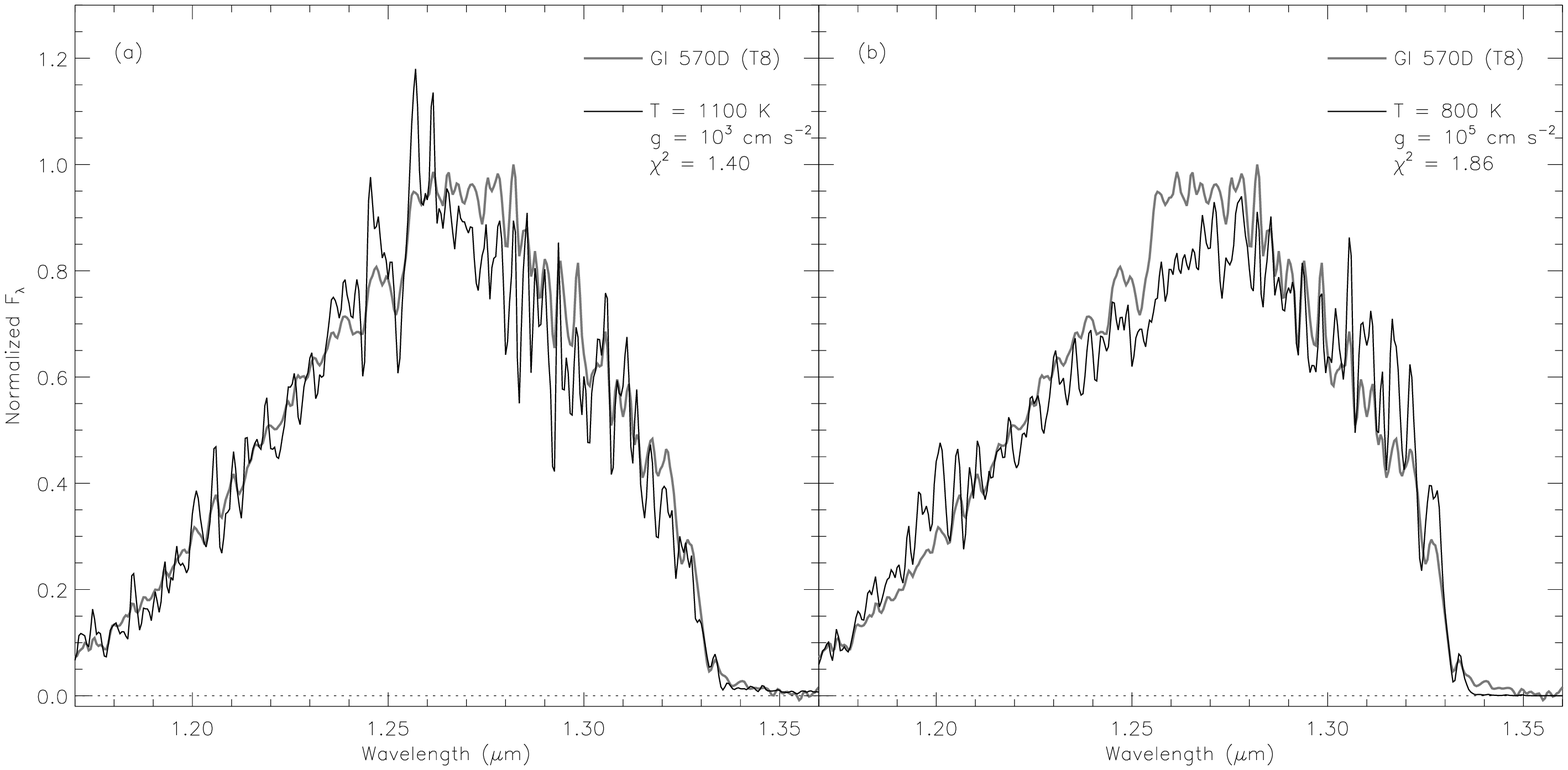}
\caption{Spectral model fits (black lines) for NIRSPEC BDSS data of the T8 companion dwarf
Gliese 570D (grey lines) to \citet{all01} models of
(a) T$_{eff}$ = 1100 K and $\log{g}$(cgs) = 3.0,
the best fit spectral model (minimum ${\chi}^2$); and (b) T$_{eff}$ = 800 K and $\log{g}$(cgs) = 5.0,
the most widely adopted empirical parameters for this object \citep{geb01}.  The best fitting model
predicts an age $<$ 1 Myr and mass $<$ 1 M$_{Jup}$, grossly inconsistent with this object's
membership in the 2--5 Gyr Gliese 570 system.}
\end{figure}

\begin{figure}
\centering
\plotone{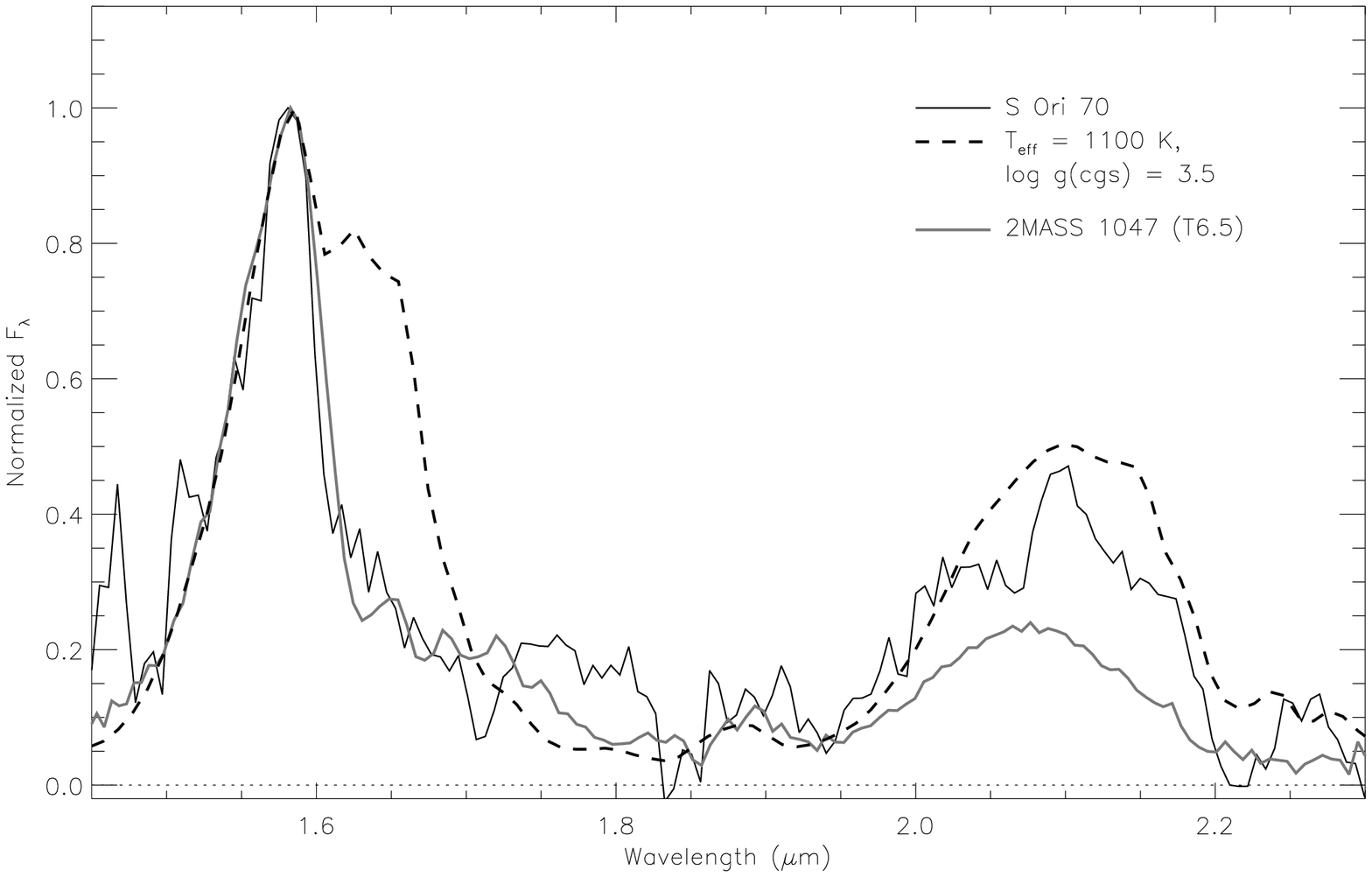}
\caption{Comparison of the $HK$ spectrum of S Ori 70 from Z02 (solid black line) to
a Gaussian smoothed T$_{eff}$ = 1100 K, $\log{g}$(cgs) = 3.5 model from \citet[dashed line]{all01}.
The model parameters are those determined by MZ03 as the best fit to
the $J$-band spectrum of this object.  Also shown are $HK$
data for the field T6.5 2MASS 1047+2124 obtained with the same instrument
\citep[grey line]{me02a}.  All spectra are
normalized at 1.57 $\micron$.}
\end{figure}

\end{document}